\documentclass[acmsmall,nonacm=true]{acmart}
\renewcommand\footnotetextcopyrightpermission[1]{} 
\makeatletter
\renewcommand\@formatdoi[1]{\ignorespaces}
\usepackage{todonotes}

\usepackage[utf8]{inputenc}

\usepackage{tikz}
\usepackage{tikz-qtree,tikz-qtree-compat}

\usepackage{xspace}
\usepackage{dirtytalk}

\usepackage{epigraph}

\usepackage{titlesec}
\titleformat{\section}
    [block]{\normalfont\bfseries\Large}{\rlap{\thesection}}{0em}
    {\hspace{.05\linewidth}\begin{minipage}[t]{.95\textwidth}}[\end{minipage}]

\usepackage{syntax}
\usepackage{amsmath}
\usepackage{amssymb}

\usepackage{todonotes}

\usepackage{listings}
\usepackage{color}
\definecolor{lightgray}{rgb}{.9,.9,.9}
\definecolor{darkgray}{rgb}{.4,.4,.4}
\definecolor{purple}{rgb}{0.65, 0.12, 0.82}
\lstdefinelanguage{TypeScript}{
  keywords={typeof, new, true, false, catch, function, return, null, catch, switch, var, if, in,
    extends, while, let, do, else, case, break, num, string, boolean, class,
    email, cssColour, type},
  ndkeywords={class, export, number, boolean, throw, implements, import, this, const},
  sensitive=false,
  comment=[l]{//},
  morecomment=[s]{/*}{*/},
  morestring=[b]',
  morestring=[b]"
}

\usepackage{amsmath}{%
    \theoremstyle{plain} 
    \theoremstyle{plain} 
}

\newcommand{\ie}{\emph{i.e.}\xspace}
\newcommand{\eg}{\emph{e.g.}\xspace}
\newcommand{\etal}{\emph{et al.}\xspace}
\newcommand{\wrt}{\emph{w.r.t}\xspace}
\newcommand{\regex}{\emph{regex}\xspace}
\newcommand{\regexes}{\emph{regexes}\xspace}
\newcommand{\safe}{\emph{SafeString}\xspace}
\newcommand{\safes}{\emph{SafeStrings}\xspace}
\newcommand{\AST}{AST\xspace}
\newcommand{\target}{\emph{TypeScript}\xspace}

\newcommand{\dsl}{\emph{DSL}\xspace}
\newcommand{\css}{\emph{css}\xspace}
\newcommand{\monoidal}{\emph{monolithic}\xspace}
\newcommand{\pcre}{\emph{PCRE2}\xspace}

\newcommand{\str}{\lstinline{string}\xspace}
\newcommand{\cssColour}{\lstinline{cssColour}\xspace}

\usepackage{lipsum}

\usepackage{cleveref}

\newlength{\emstr}
\newcommand{\boldpara}[1]{%
         \smallskip%
          \par\noindent\textbf{\textit{#1}}\hspace{\emstr}
}%

\begin{document}

\title[SafeStrings]{SafeStrings \\ Representing Strings as Structured Data}

\author{David Kelly} 
\affiliation{\institution{University College London}}
\email{david.kelly.15@ucl.ac.uk}

\author{Mark Marron}
\affiliation{\institution{Microsoft Research}}
\email{marron@microsoft.com}
\author{David Clark}
\affiliation{\institution{University College London}}
\email{david.clark@ucl.ac.uk}
\author{Earl T. Barr}
\affiliation{\institution{University College London}}
\email{e.barr@ucl.ac.uk}

\renewcommand{\shortauthors}{Kelly et al.}

\begin{abstract}
Strings are ubiquitous in code. Not all strings are created equal, 
some contain structure that makes them incompatible with other strings. 
CSS units are an obvious example. 
Worse, type checkers cannot see this structure: this is the latent structure problem.
We introduce \safes to solve this problem and expose latent structure in strings. 
Once visible, operations can leverage this structure to efficiently manipulate it; 
further, \safes permit the establishment of closure properties. \safes
harness the subtyping and inheritance mechanics of their host language to create 
a natural hierarchy of string subtypes. \safes define an elegant programming model 
over strings: the front end use of a \safe is clear and uncluttered, with complexity confined 
inside the definition of a particular \safe. They are lightweight, 
language-agnostic and deployable, as we demonstrate by implementing 
\safes in \target. \safes reduce the surface area for cross-site scripting, 
argument selection defects, and they can facilitate fuzzing and analysis.
\end{abstract}

\maketitle
\section{Introduction}\label{sec:intro}
Strings are the great dumping ground of programming.
Their meagre structural
obligations mean that it is easy to pour information into them.
Developers use them to store semi-structured, even structured, data.
Thomas Edison (paraphrasing Reynolds) captured the reason: ``
There is no expedient to which a man will not resort to avoid the real
labor of thinking.''
Take colour assignment in \css.
As flat strings they take the form \lstinline+'#XXX'+ or \lstinline+'#XXXXXX'+,
but they implicitly encode three integer values for \lstinline+rgb+. Colour operations are
naturally arithmetic, but as strings they are cumbersome and error prone for
developers to use.
Simple operations are difficult: logically correct equality \lstinline+'#000' == '#000000'+ is false.  
Incorrect assignments are common: \lstinline+x = '#F65T00'+ can only be caught by
the developer.
We call this problem the \emph{latent structure problem}. This is a widespread problem.
Finding errors of this sort is extremely difficult, and
programmers care about these errors.

The \target community has seen a great deal
of discussion about \emph{string literals} \cite{stringlit} and
\regex-validated strings~\cite{regexlit}. Regex here refers to PCRE2 style expressions, which
augment regular expression with backtracking, grouping, and lookahead.
A quick internet search for
``validated strings'' returns over ten million hits discussing solutions to the
problem\footnote{Google 29 March 2019.}.
The latent structure problem bedevils debugging; it also introduces security vulnerabilities: format string attacks and
SQL injection attacks are well known examples \cite{daglib}. Handling strings
properly does not just give greater \emph{type} safety, it gives greater safety.

Current type checkers cannot tackle the latent structure problem
because they cannot reach past \str to the underlying structure; 
many problems in string theory
are undecidable, making many analyses either dependent on heuristics or very expensive.
Strings seem an easy method for representing structured information: instead, they are a great way to
\emph{de}-structure information. Latent structure encoded in strings is \emph{stringly typed}.
The consequence is that programs pass around and manipulate richly-structured strings as if they had little or no structure.

Industry has been working to solve the latent structure problem. To date, the dominant
approach is string validation by adding language support for
checking that a string is what it claims to be. This is the tack
taken in a recent pull request for \regex-validated strings in \target
\cite{regexlit}. This request has been open since 22 January 2016 and has had
over 50 detailed comments.
It uses regular expressions to check that a string has the desired ``shape''.
A raw string is checked for certain properties and this information is used to instantiate
a string \emph{subtype}. This catches the \css colour error:
\lstinline+/^#([0-9a-f]{3}|[0-9a-f]{6})/+ does not accept
\lstinline+'#F65T00'+.
The \pcre is lifted to the level of types, becoming part of the type
declaration (\autoref{lst:emailregex}). This is clearly powerful,
but it has not been added to the language, despite the enthusiasm for the idea evident in the
request and despite the working code. 

It is programming folklore that regular expressions add extra problems without
resolving the old ones. In type declarations, this is even more true: a \css
colour has a reasonably simple structure, but a regular expression for an
email address is not for the faint-hearted:

\begin{lstlisting}[mathescape,framerule=0pt,caption={Email \pcre type as
    suggested in \cite{regexlit}.},label={lst:emailregex}]
  type Email = /^[-a-z0-9~!$\mbox{\textdollar}$%^&*_=+}{\?]+(\.[-a-z0-9~!$\mbox{\textdollar}$%^&*_=+}{\?]+)*@([a-z0-9_][-a-z0-9_]*(\.[-a-z0-9_]+[a-z][a-z])|([0-9]{1,3}\.[0-9]{1,3}\.[0-9]{1,3}\.[0-9]{1,3}))(:[0-9]{1,5})?$\mbox{\textdollar}$/i];
\end{lstlisting}
\noindent
Clearly, regular expression validated strings are difficult to write and difficult to
maintain. They do not admit of any easy notion of subtyping, even though it is
reasonable to say that a \lstinline+gmail+ address should be a subtype of
\lstinline+email+. Regex-validated strings are essentially immutable and the
question of what type results when applying operations to them is not even
addressed. Any change in a string necessitates rechecking the entire string, changing the \emph{red} element of a \css colour for instance. This is expensive and wasteful.

We introduce \safes to solve the latent structure problem.
Like the \pcre validation approaches, \safes expose the latent structure of a
string to the type checker. This is where the similarity ends.
While \regex-validated strings check, find structure, and then
throw the memory of that structure away, \safes retain and manipulate the
structure, taking what is latent and making it manifest.
\safes allow operations closed over the type: an operation can be guaranteed to
return the same type of \safe (\autoref{sec:liquid}). As they store an image of the string's
structure, \safes do not need to recheck the entire string when performing updates, unlike
with \regex-validated strings.

\safes make the latent structure of the string available to the type checker
(\autoref{sec:safestrings}). \safes represent \css
colour string (\autoref{sec:casestudies}) internally as
\lstinline[mathescape]+(Hash (Red $\mathbb{N}$) (Green $\mathbb{N}$) (Blue $\mathbb{N}$))+.
A special operation \lstinline+cast()+ allows us to reconstruct the original
string: \lstinline{cast() : string = '#' + red.toHex() + green.toHex() + blue.toHex();}.
\emph{Composable} recognisers mediate habitation of the structure, \eg each structural
element has its own independent parser and the string is tested by putting them
together in the correct order:
\lstinline+cssColour = parseHash.then(parseRed.then(parseGreen.then(parseBlue)));+.
We reject \lstinline+'#F65T00'+ as malformed, but we can also correctly allow
the equality
\lstinline+x : CssColour  = '#000'; y : CssColour = '#000000' ; x == y = true;+ as the \emph{structure} of both
strings is the same (\autoref{subsec:equality}). Of course, raw string equality is still possible. 
\safes model subtyping elegantly (\autoref{subsec:subtyping}). The internal
representation of the string 
as a structure means subtyping comes naturally from overriding structural elements. 
In the case of email addresses, a generic recogniser for the
domain part of the address is replaced with one especially tailored for `gmail'.
There is no need to recheck the entire string (\autoref{subsec:op}).

Capturing information in a type is familiar in functional
programming, where it is common to define a \emph{monad} or \emph{applicative
  functor} over a structure, then define functions over it. This
pattern is ultimately derived from categorical thinking: the
recognition that certain types of structure (functors) and functions (evaluation)
define algebras. \safes realise that the latent structure problem
can be solved in strings by modelling them as functor algebras (\cref{sec:safestrings}).

To evaluate \safes we consider three case studies of useful string types:
filepaths (\autoref{subsec:filepath})
css strings (\autoref{subsec:css}) and 
email strings (\autoref{subsec:email}),
in \target. \safes are a lightweight annotation system that exploits the structure available in
string literals. \safes support an elegant and natural programming model over strings by
pushing the complexity away from the programmer.
This is in contrast to \regex-validated strings, where the complexity sits
directly in the type signature, for limited gains.
We demonstrate a possible concrete
syntax for \target which we implement via a simple preprocessing stage (\autoref{sec:deploy}). The
preprocess stage adds only a sprinkle of syntactic sugar: \safes are immediately deployable.
\safes do not require any special language mechanisms beyond the ability to
represent structured data and to define recognisers over this data.

Exposing latent structure necessarily exposes complexity. Deciding \emph{who}
deals with that complexity and \emph{when} is vital for the usability of any
programming model. \safes provide a simple and spare user-facing programming
model; they move the intrinsic complexity away from front-end developers to
library authors.

Our principle contributions are:

\begin{itemize}

    \item We identify the \emph{latent structure problem}, the gap 
      between a language's representation of data and the fine-grained 
		way programs actually use and structure that data;

    \item To solve the latent structure problem, we introduce \safes, a
	    lightweight, language-agnostic, and immediately deployable
		programming model (\autoref{sec:safestrings});

    \item We show how \safe faciliate the definition and verification of type
	    safe operations over strings (\autoref{subsec:closure}); and

    \item We present a realisation of \safes for TypeScript
	    (\autoref{sec:implementation}).

\end{itemize}

The latent structure problem is not limited to strings. Lists, for example, are
also likely to have additional latent structure. The prevalence of strings in
programming means that the problem is often most acutely felt in that domain.
\safes are integratable with the notion of value literals, creating
\emph{SafeLiterals}, a uniform and easily extensible mechanism for introducing
typed value literals. This extension provides a simple solution to the
proliferation of \emph{ad-hoc} object parsers and under-constrained JSON encodings
to represent literal values in \target and other programming languages. All
artefacts are available at \url{anonymised.repo.to.do}.

\section{SafeStrings}\label{sec:safestrings}
\safes address the latent structure problem for strings. 
Strings used in programming, such as phone numbers,
filepaths and zipcodes, tend to be highly structured. \safes harness this latent
structure and surface it as required. This surfacing brings greater type safety,
greater control over mutability, and powerful subtyping capabilities.

In what follows, we use the term \emph{regular expression}
for expressions over the regular operators that capture only 
the regular languages.
We use \emph{PCRE2} to mean string matching expressions now in widespread use
in industrial programming languages.
These have regular expression-like syntax, but are strictly more expressive, due
to the presence of backtracking, grouping, and lookahead. The pull request for
\regex-validated strings \cite{regexlit} uses JavaScript-style expressions.
These are \emph{PCRE2}.
We use \str{} to refer to the string \emph{type} of
a language and \emph{string} or \emph{raw string} to mean a string in the
sense of a list or array of characters, without reference to its exact
representation (\eg object or null-terminated list) in a
given language.
We assume that raw strings are finite to simplify the presentation\footnote{A
\emph{stream} might be modelled as a \safe if finite prefixes of the stream had
an encodable structure. We leave this for future work.}.

Informally, \safes combine a string, a
grammar and a recogniser. A \safe is the set of all strings $s$ such that they are in the
language of the given grammar, \ie \safe = $\{s \mid s \in L(G)\}$,
where $L(G)$ is the language of the grammar. One can readily see that a \safe
expresses a subset of strings. 
The recogniser is a parser (\autoref{sec:implementation}).
In essence, a \safe is a \dsl (Domain Specific Language) embedded in the native
\str type of a language and specialising it in a principled fashion.
A \safe recogniser mediates this embedding: only raw strings accepted by the parser are encoded into \safes. 

Operations take place inside the \dsl rather than in the less structured
world of strings. The original, \emph{raw} string can be recovered
through a casting mechanism that has rules for recreating the original string
from the structure.

This is a lot of additional work to make strings safe. It is necessary because
it is difficult to extract structure from strings due to their form. They are
just the free monoid over an alphabet, $\sigma$, with 
concatenation as the binary operation, $\cdot$, and the empty string,
$\epsilon$, as the unit. This provides the type checker with relatively little
information. Any combination of characters makes a valid string. No other common
data types are similarly unconstrained. Finding and encoding more structure enables the type
checker to make stronger static guarantees about the dynamic behaviour of a program.

An email address \lstinline+safe@mail.com+ viewed solely as a \str is uninteresting,
but represented as a \safe provides a much greater amount of information
available for analysis.
An email string, for example, could be represented by a structure such as
\begin{lstlisting}[mathescape]
  (At '@' (Name string$_1$) (InvariantDot '.' (Left string$_2$) (Right string$_3$)))
\end{lstlisting}

The structure is an abstract syntax
tree (\AST). Each substring needs to be checked. It is not sufficient for a
string to merely have the scaffold (\ie `@' and `.') of an email address to
qualify as an email string. Each element of the structure has to have an associated 
recogniser.

Using a single PCRE2, the recogniser for an email address is, as we have seen
(\autoref{sec:intro}) formidable. Given the power to compose and order recognisers
however, it is a much simpler proposition:

\begin{lstlisting}[framerule=0pt]
  InvariantAt  = /@/
  Name         = /[0-9a-zA-Z]+/
  InvariantDot = /\./ 
  Left         = /[0-9-a-zA-Z-]+/ 
  Right        = /[0-9-a-zA-Z-.]+/
\end{lstlisting}
This presentation risks losing the string itself, and a \safe should behave like
a string when required: we therefore also require a \lstinline+cast() : email -> string+
function that `reconstructs' a \str from the representation. Its 
exact definition depends on the details of the representation.

\begin{definition}
  A \safe is a 4-tuple $\langle G, R, \phi, c \rangle$ where
\begin{align}
	G : &\ Grammar \\
	R : &\ G \rightarrow \text{string} \rightarrow \mathit{AST}_{\bot} \\
	\phi : &\ \mathit{AST}
		\rightarrow \mathit{AST}^\textsc{PL} \\
	\alpha : &\ \mathit{AST}^\textsc{PL} \rightarrow \text{string}
\end{align}
$G$ is the grammar. $R$ is a recogniser, a (partial) function that takes a grammar and an input
string, and returns \AST$_{\bot}$, lifted to include $\bot$, capturing the
possibility of failure. $\phi$ takes the \AST
and maps it to \emph{AST}$^\textsc{PL}$, a concrete realisation of the \AST in
the language \textsc{PL}. \textsc{PL} may be a programming language (\ie target)
or a family of languages (e.g. \textsc{LLVM IR}).
$\alpha$ is the \lstinline+cast()+ operation, taking
\AST$^\textsc{PL}$ back to a raw string.
\end{definition}

The fundamental insight for \safes comes with $\phi$: previous work
on validated string types performs the check $R$ but does not convert and store
the \AST. The hard work of verification is discarded rather than kept and
utilised. The handling of the error condition, $\bot$, in $R$ is implementation
specific; one can just fail, or return information about the error (\ie a
parse error). $\mathit{AST}^\textsc{PL}$ 
might take the form of an object or an abstract data type. 
We are assuming, of course,
that a string has \emph{some} structure. A string with no discernible grammar (\ie a randomly generated string)
is not a good candidate to be a \safe.
At its simplest, the grammar conforms to a regular language, but is not
limited to such: \safes easily embed context-free grammars and beyond.
The CFGs (context-free grammars) of likely use cases for \safes are, in general,
uncomplicated.

At its most general, the structure of a \safe is an \emph{F-algebra} $(A, \alpha)$. Let $F$ be
an endofunctor on a category $C$, $F : C \rightarrow C$, then $A$ is the carrier
set and $\alpha$ a morphism of the form $F(A) \rightarrow A$. $F$ corresponds to
the $Grammar$, whereas $A$ is the set of strings. The `interpretation'
morphism $F(A) \rightarrow A$ can now seen to be \lstinline+cast()+ of
\autoref{sec:intro}, specialised in that case to strings.
In general, the habitation of a \safe is conditioned solely on its syntactic
well-formedness.  Whether the resulting \safe is \emph{meaningful} depends on
the sophistication of the recogniser: a syntactically correct email may identify
no recipient.

An \emph{invariant} in a \safe is a sub-string of the raw string that always
occupies the same position \wrt the other elements of the string. 
Both the ``@'' symbol and the \emph{first} dot ``.'' of an email address are
invariants. They dictate the scaffold of the string around which the other
elements (substrings) are disposed.  Consider  an `\emph{inner-r}' string. A suitable representing structure
would be:
\begin{lstlisting}[framerule=0pt,mathescape]
  (InnerR (InvariantR 'r') (Left str$_1$) (Right str$_2$))
\end{lstlisting}
\noindent
\lstinline+'r'+ is an invariant element in an \emph{inner-r} string and need not be
specifically encoded in the representation, \ie the representation could be
\lstinline[mathescape]+(InnerR (Left str$_1$) (Right str$_2$))+.
The \lstinline{'r'} is
trivially recoverable by defining the \lstinline+cast()+ method as
\lstinline[mathescape=true]{cast = str$_1$.cast() + 'r' + str$_2$.cast()}.

The space requirements of \safes depends on the size of \AST$^{pl}$.
In general, a \safe requires more space than a raw string.
A well designed encoding, however, can make the \safe representation more
space efficient than that of the raw string. Consider the archetypal context-free language of
equal numbers of $a$s and $b$s, $a^{n}b^{n}$. Such a string can be arbitrarily
long, but the \safe need only record the structure and the recogniser as a PCRE2
(\autoref{lst:asandbs}).
A \safe  is not limited solely to PCRE2 for membership tests. We can write
recognisers that also count string length or perform other operations that are not explicitly linked
with membership. The \safe \lstinline+EqualAandB+ also counts the number of inputs.

\begin{lstlisting}[float=t,caption={A \safe of balanced letters.}, label={lst:asandbs}]
  class EqualAandB = {
    count : num
    cast() = 'a'.repeat(count) + 'b'.repeat(count);
  }
\end{lstlisting}

Methods for string safety, such as type aliasing or \regex-validated strings
(\autoref{sec:related}), are special, degenerate cases of \safes. We call these
special cases \monoidal \safes.
\emph{Monolithic} \safes invoke $R$ whenever they want to expose and access the
internal structure. It does not store the \AST.
They only need the recogniser $R$, and do not define $\phi$ and $\alpha$.
In effect, a \monoidal \safe matches the entire string and places that string
inside a wrapper. This wrapper exposes the string to the type checker without
exposing the latent structure of the string:

\begin{lstlisting}
  class Monolithic {
    // recogniser check on input
    raw : 'no_recorded_structure'
  }
\end{lstlisting}

A \monoidal \safe is the same as Fowler's wrapper type (\autoref{sec:related})
and is isomorphic to \regex-validated string literals. This can easily be
seen by noting that the only
salient different between \regex validated types and \monoidal \safes is the
method by which a new type is created, \ie either by privileged keyword or via a
native record. As we shall see in~\autoref{sec:liquid}, \monoidal \safes, while
easy to implement, lack much of the flexibility one gets from treating them with
a more complex representation.
We briefly explore the implementation of \monoidal \safes and \safes over
relatively simple structures as captured by PCRE2, showing how
they subsume earlier work (\autoref{sec:related}).
More interesting \safes involve grammars with multiple
constructors and recognisers that match elements to constructors as well as
accepting the input string.

\safes permit the efficient definition of function sensitive operators.
Even a very simple regular language need not be treated as a \monoidal \safe.
The regular expression  $/^.*r.*$ recognises structure and therefore can be made
modular. For `\emph{inner-r}' strings, \safes has isolated the \lstinline+Left+ and \lstinline+Right+ substrings 
without losing their relation to the main structural element.
 We now examine the
benefits \safes provide by exposing the latent structure of a string.

\section{Subtyping and Operations over \safes}\label{sec:liquid}
Type systems, at their most general, capture program invariants and use this information to provide
feedback to a user. 
This feedback is most often in the form of a \emph{type error}
which typically results in a compilation failure or cessation of interpretation.
Type systems are usually classified by whether they are static or dynamic.

One of the main advantages of
static type systems is they capture an abstraction of the program's dynamic
invariants at compile time. This comes at the necessary cost of over-approximation.
Classical type systems especially
only capture a relatively coarse image of the program's behaviour.
For example, most type systems 
capture the fact that a variable $i$ is an integer\footnote{An obvious
  counterexample is \target, which has a general \emph{number} type. }.
They cannot however model the fact that an integer is even, natural number.
Dependent types \cite{Martin-Lof} are sufficiently expressive to
capture this kind of invariant at the type level. Dependent types come with a
large annotation and cognitive burden on the programmer (\autoref{sec:related}).

Liquid types are a decidable subset of dependent types so can strip away
much of the associated annotation complexity. This decidability means
that constraints can be inferred statically without proofs required from the
user, in much the same manner that `normal' types themselves can be inferred in a
\emph{Hindley-Milner} type system. Liquid types are necessarily less expressive
than full dependent types, but the decidability makes them more useful for
regular programming.

\safes capture much of the expressive power of liquid types and apply it to
strings. 
Given that much of string theory is undecidable
\cite{ganesh2011decidable}, it is inherently difficult to have a
decidable fragment of string theory suitable for instantiating liquid types.
Many useful fragments of string theory are decidable however \cite{chen2017decidable} and progress has
been made on solvers for string theory~\cite{berzish2017z3str3}. These
approaches to string safety are complementary to our approach, and could be integrated into a
language's type checker.

The problem is compounded when we consider operations specialised over
strings. As is well known, the concatenation of two strings is a string, but
the concatenation of two email strings is very unlikely to be a well-formed email string.
It is,
however, still a \str, so the operation is closed under its supertype (\autoref{subsec:closure}).
Exposing this information to a type checker allows
us to have liquid typing for specialised strings. Moreover, it provides the
opportunity to specialise \emph{operations} over \safes. Thus \safe
concatenation can be written in two forms, one which can be checked to still be
within its type, and one which automatically is upcast to \str.

We discuss the relationship between \safes and liquid types \wrt operations over
strings in the remainder of this section. We start by discussing subtyping for
strings and \safes.

\subsection{Subtyping}\label{subsec:subtyping}

Even though strings are often represented by objects in many languages, they do
not easily admit of a principled subtype relation, especially if that subtype of
string has validation conditions (\autoref{fig:subtyperegex}). An inheritance
relation in most \emph{OOP} languages is created by extending and altering the
structure of an object. Treating the string as a structure, as in \safes, allows
the same subtyping and inheritance mechanisms to be brought into play. It also
simplifies the problem of validation (\autoref{fig:subtypingTSC}).

We may categorise subtyping broadly between nominal and structural: nominal
subtyping means that one type is the subtype of another if so declared (\ie a
named relationship is sufficient), whereas
structural subtyping considers two types compatible if they share corresponding
elements, regardless of names. \emph{Java} is a notable example of a language with nominal subtyping,
whereas \target has structural subtyping.

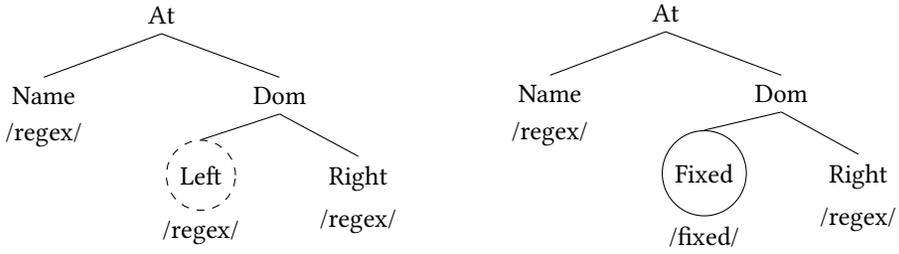
\begin{figure}
  \begin{minipage}[l]{0.4\textwidth}
    \begin{tikzpicture}
      \tikzset{level 1+/.style={sibling distance=2\baselineskip}}
      \Tree
      [.At
        [. \node[label={below:/regex/}]{Name}; ]
        [.Dom
          [. \node[draw,circle,dashed,label={below:/regex/}]{Left}; ]
          [. \node[label={below:/regex/}]{Right}; ]
        ]
      ]
    \end{tikzpicture}
  \end{minipage}
  \hspace{1cm}
  \begin{minipage}[r]{0.4\textwidth}
    \begin{tikzpicture}
      \tikzset{level 1+/.style={sibling distance=2\baselineskip}}
      \Tree
      [.At
        [. \node[label={below:/regex/}]{Name}; ]
        [.Dom
          [. \node[draw,circle,label={below:/fixed/}]{Fixed}; ]
          [. \node[label={below:/regex/}]{Right}; ]
        ]
      ]
    \end{tikzpicture}
  \end{minipage}
\caption{Subtyping for a generic email \safe string to specialised email string.
As each node of the tree is associated with a sub-recogniser (here a regular
expression), it is necessary only to override the appropriate recogniser. As the
structure as not changed, it is not necessary to recheck the other nodes.}%
\label{fig:structuralsubtyping}
\end{figure}

In languages that have subtyping, it is reasonable to ask how
subtyping might work on \safes. Using \monoidal \safes does not provide the
necessary structure to satisfactorily answer this question.
One method is to ask if one language is included within another.
The inclusion problem for regular languages is PSPACE-complete 
\cite{meyer1972equivalence}), although a subset of regular language inclusion problems can be decided in polynomial
time \cite{hovland2012inclusion}. In particular, inclusion for $RE^{\leq k}$ is
in $P$, where  $RE^{\leq k}$ is the class of regular expressions where each
symbol appears at most $k$ times. The complexity is still poor, on the order of
$n^k$, with $n$ being the length of the regular expression.
In general, however, the problem of language inclusion
for simple regular expressions is intractable \cite{Martens04}.
Moreover, these bounds hold only for regular
languages, they do not hold for the complexity of \regexes.

There are various methods to simulate a subtype relation without resorting to
language inclusion.~\autoref{fig:subtyperegex} shows one such solution as suggested
in the pull request for \regex string literals in \target~\cite{regexlit}.

\begin{lstlisting}[mathescape,
caption={A possible, but expensive, solution to the problem of subtyping for
  regex-defined string literals as found in the original pull request \cite{stringlit}. This method would also work for one-field
  \safes. In addition to obvious efficiency problems due to redundancy, it
  suffers from readability and maintainability problems.},%
label={fig:subtyperegex}%
]
let email : Email;
let gmail : Gmail;

type Gmail = Email & /^[-a-z0-9~!$\mbox{\textdollar}$%^&*_=+}{\?]+(\.[-a-z0-9~!$\mbox{\textdollar}$%^&*_=+}{\?]+)*@gmail\.com$\mbox{\textdollar}$/i;
gmail = email;  // correct
\end{lstlisting}

This solution is less than ideal: it does a complete check twice. First it acertains
whether the input string is of type \lstinline+Email+ and then looks
again to assert membership in the second \regex, which introduces the specialisation. 
This is clearly of limited extensibility and maintainability. Moreover,
given the backtracking and lookahead abilities in \regex{}es, this could be a
very expensive operation.

\safes address the problem of subtyping via manipulation of the structure (\autoref{fig:structuralsubtyping}). 
An \lstinline+email+ \safe might have a generic
structure similar to \lstinline+{name, at, domain}+ where `at' is an
invariant, a necessary requirement for string membership
(\autoref{sec:safestrings}).
It is reasonable to accept a \lstinline+gmail+ \str to be
a specialisation of an email. This requires that the domain contain the substring `gmail'
in the appropriate position. Given a set of composable parsers,
a \lstinline+gmail+ subtype of an email
address is a specialisation of one, or more, of those sub-parsers. In
\autoref{sec:casestudies}, parser combinators are used. These
are ideally suited for this role due to there composability.
We note however that this is not an essential requirement (\autoref{sec:related}).

\begin{lstlisting}[float=t,
caption={Subtyping email \safes in pseudocode similar to \target. Subtyping can
  be done in the usual fashion, by overriding and extending.},%
label={fig:subtypingTSC}%
]
  export class Email {
      user : Parser = parseName;
      dom  : Parser = parseGenericDom;

      email : Parser = parseName.then(parseAt).then(parseGenericDom)
  }

  Gmail extends Email {
      static dom  : ParserGmailDom;
  }

  MoreGmail extends Gmail {
    static extra : Parser = parseMore;
    email = super.email.then(extra)
  }
\end{lstlisting}

An inheritance relation can thus be formed naturally either by changing the
rules for membership in structural elements, or by adding additional parts to
the structure.

\subsection{Equality}\label{subsec:equality}
Equality, broadly speaking, carries the same dichotomy as subtyping: nominal or
structural. The details are again language dependent. 
Structural equality for \safes asserts that
\lstinline+foo : string = "foo"+ and \lstinline+ foo' : Foo = "foo"+ evaluate to \lstinline+true+
despite the fact that they have different types. This is reasonable behaviour, as \lstinline+Foo+
is presumably a subtype of \lstinline+string+. A purely nominal approach however would evaluate to
\lstinline+false+, as the annotated types of the two declarations differ. 

Ideally one would want to have access to both behaviours. The former reflects the intuition that one
is dealing only with strings \emph{unless otherwise required}, the latter when
we require the \str to have a checked
structure.

\target already has access to two forms of equality, inherited
from JavaScript `==' and `==='. This gives us an easy way to model the two
desired behaviours. Each
\safe therefore requires a definition of a method, \lstinline+eq+, in two flavours,
\emph{strict} and \emph{weak}. Weak performs a cast of a \safe to its string
representation, and then performs the equality test. \emph{strict} checks
the type instance first before progressing to checking the
contents.~\autoref{sec:casestudies} gives more detail about defining these
important functions.

\subsection{Typesafe Operations}\label{subsec:op}

One of the main advantages of liquid types is that they can confirm \emph{pre}
and \emph{post} conditions overs primitive types under operations. For example,
an integer refined to be \emph{even} only can be shown to be even after an
operation such as multiply by two. Perhaps more importantly, it can also show
when an assertion fails. \safes bring some of this discriminatory power into
strings.

A \safe, as we have seen, has an internal representation. This representation is
sequence or tree-like. Operations over a \safe are best viewed as tree surgery
over one or more nodes. It is not usually necessary to operate on the
\emph{entire} tree.
We have already seen this idea in \autoref{sec:intro} in the
discussion of \cssColour strings an example where it is necessary to operate
over the entire tree, but doing so results in a rational, type safe function (\autoref{fig:blend}).

\begin{lstlisting}[float=t,label={fig:blend},%
caption={Blending two \cssColour strings together in guaranteed to produce a
  well-formed \cssColour string. This operation is closed over its type due to
  the fact that a \cssColour string is structurally an integer, and addition has
  the required properties.}%
]
  // the addition ``+'' operation for \cssColour strings
  blend(s1 s2 : CssColour) : CssColour {
    red = safeAdd(s1.red + s2.red) // assume that safeAdd correctly handles overflow 
    // other components the same
    col = new CssColour(red, green, blue)
    return col
  }
\end{lstlisting}

In general however, one need only edit one or two nodes. This means that, in the
event that we do not have guaranteed closure, we need only recheck the substring
that has been altered, without having to recheck the entire string. To return to
one of our running examples, email strings, we might want to generate
pseudo-random addresses via string concatenation. This procedure could be useful
in, \eg test cases generation or fuzzing. Naturally, we want this concatenation
function ``+'' to behave in as safe a manner as possible, and providing as such
information to the type checker as is possible.

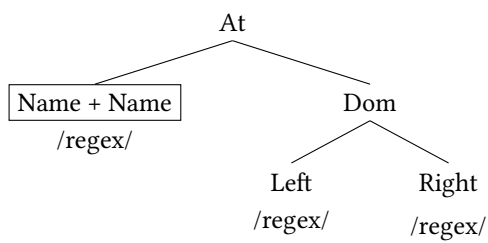
\begin{figure}
\begin{tikzpicture}
  \tikzset{level 1+/.style={sibling distance=2\baselineskip}}
  \Tree
  [.At
    [. \node[draw,label={below:/regex/}]{Name + Name}; ]
    [.Dom
      [. \node[label={below:/regex/}]{Left}; ]
      [. \node[label={below:/regex/}]{Right}; ]
    ]
  ]
\end{tikzpicture}
\caption{Editing individual nodes in the tree representation of a syntactically valid email address.
    Email addresses concatenation defined only over the name field. Depending on the
    \regex, this can be resolved purely statically (\ie the concatenation of the
    \regex is isomorphic to the original) or with a runtime check.}%
\label{fig:edit}
\end{figure}

Depending on the precise recogniser associated with \emph{name}, the ``+'' can be
checked either statically or dynamically for closure. Given a \emph{name}
recogniser of \lstinline{/[a-zA-Z0-9]+/} for example, it can be readily seen that
concatenating two strings matching this pattern are still within the pattern. A
real email address name field is more complicated of course.
Regular string concatenation can still be had via \lstinline{cast()} (\autoref{lst:concat}).

\begin{lstlisting}[float=t,caption={Typesafe email concatenation and normal string
    concatenation in \target. The user can overload the operation by choosing to
  constrain the return type.},label={lst:concat}]
    let email1 : Email = 'foo@bar.com' 
    let email2 : Email = 'bax@bar.com'

    // overload + to perform safe email concatenation over names
    'foo@bar.com' + 'bax@bar.com' // constrained to Email.
    => 'foobax@bar.com'

    // raw string concatenation with return type string. Not a valid email address.
    'foo@bar.com' + 'bax@bar.com' 
    => 'foo@bar.combax@bar.com'
\end{lstlisting}

Anther common operation over a string is slicing, \ie extracting a substring.
The key to recognising the power of slicing in \safes is that we
normally are looking to extract a \emph{particular} substring from a string, and
not just a random slice. These particular substrings are likely to be
represented in the structure as individual elements or the composition of
individual elements. Slicing then becomes merely projection.
\autoref{lst:badslice} shows one way to perform this in \target.
Given a \regex-validated string, one still needs to write \lstinline+email.split('@')[0]+. With a
\safe however, one simply extracts the named field, \ie \lstinline+email.name()+, an action that 
has no additional runtime cost of extra function calls or string manipulation. Moreover, it
cannot fail silently as in the manner of \autoref{lst:badslice}. A user desiring
free slicing over a \safe need only manipulate the raw strings as opposed to its
representation, \ie take a slice from the \lstinline+cast()+ of the \safe.
We say about type safe string mutation when discussing various
\safes in \autoref{sec:casestudies}.

\begin{lstlisting}[float=t,label={lst:badslice}, caption={Slicing in TypeScript}]
"someone@email.com".split('@')[0] // OK: returns 'someone'

"someoneemail.com".split('@')[0] // Bad: returns 'someoneemail.com' without warning
\end{lstlisting}

Finally, we note a use of \safes that is not directly related to their form or
mutability. The problem of \emph{argument selection defects} is well known.
This is concerned with function parameters of the same type being used in the
incorrect order. \cite{Rice:2017} give the motivating example found
in \Cref{lst:argdefect}. A developer has mistakenly swapped the parameters in 
\lstinline+getUser+.  Appropriately typing these strings as \safes would prevent this
error from having occurred, meaning that the much more expensive testing process
as detailed in Rice \etal is not required in this instance. While \safes do not
eliminate the problem (two \safes of the same type could still be
swapped), the surface area for such errors is now much smaller. 

\begin{lstlisting}[float=t,caption={Argument selection defects are compounded when the
    parameters are all typed as \str. \safes reduce the instance of such defects
  by giving meaningful \em{type} information, rather than relying on bi-modal
  information channels.},label={lst:argdefect}]
private User getUser(String companyId, String userId) { ... }

public void doSomethingWithUser(String companyId, String userId) {
  User user = getUser(userId, companyId)
}
\end{lstlisting}

\subsection{\safes, Solvers, and their Applications}

The question of the decidability of string theory, \ie the problem of automatically solving string constraints, has seen a
recent renewal of interest in the research community~\cite{Abdulla2017, Abdulla2014}. This is doubtless due to the
recognition of the importance of analysing string manipulating programs. A large amount of recent work has focused on the development
of practical string solvers. The list of solvers that now handle at least a part of string theory includes, but is not
limited to, Z3-str~\cite{Zheng:z3}, CVC4~\cite{Liang:2014} and Stranger~\cite{Yu:2014}.  The primary use case for these solvers
is in symbolic analysis (King~\cite{King:1976}, Cadar \etal~\cite{Cadar:2006}). A symbolic analysis systematically explores
executions in a program and collects symbolic path constraints. A symbolic executor passes these constraints to a solver 
to determine which program locations to continue exploring. For this to be a practical approach with strings, the constraint
language must precisely model strings in the language of the program under test. 

Solvers are frequently used as the back end when implementing liquid types. Perhaps the most well known
implementation is that of \emph{Liquid Haskell} (Vazou \etal~\cite{Vazou:2014}), but other targetted
languages are ML (Xi and Pfenning~\cite{Xi:1999} and F$\sharp$ (Swamy \etal~\cite{Swamy:2011}). The observant
will notice that these are all statically typed, functional languages. Attempts to bring refinement types
to so-called `scripting' languages (such as \target) have proven difficult
(Chugh \etal~\cite{Chugh:2012}). This difficulty is largely due to the
interaction between higher-order functions and imperative updates.
Vekris \etal~\cite{Vekris:2016} develop a refinement type system for \target that is
primarily concerned with addressing these issues.

\safes are complementary
to this SMT-based approach. \safes hold much interesting promise for symbolic
execution in that they can reduce the constraint space over a program. The
encoding of the \safe can be passed to the SMT-solver allowing it to form
constraints over a much more detailed representation that just \str. \safes also
reduce the complexity of fuzzing string based programs. As the structure of a
\safe is explicit, and a program has type checked, then, at any particular
program point, it is known that a \safe has the structure specified. 
A fuzzer or test generator can use this informatoin to ease the problem of
finding malformed inputs and writing unit tests.

\subsection{Closure Properties in \safes.} \label{subsec:closure}

Under \safes, a even simple expression language has some surprising
and enviable properties.

\begin{lstlisting}[framerule=0pt]
  <expression> ::=
    <const> -> number
    <op>    -> expression expression
\end{lstlisting}

As a \safes grammar, any projection from this structure is
a well-formed expression. Consider the encoding of the string ``3 * 4 + 5''
as \lstinline+(Add (Mult (Const 3) (Const 4)) (Const 5))+.
The available projections from this algebra are the string itself and
\lstinline+{(Mult (Const 3) (Const 4))+, or ``3 * 4''. We say a \safe for this
  expression language is closed
under projection. Such a useful state of affairs does not obtain in general. The
projection of the \emph{name} from a email \safe is not itself an email address.

Going in the other direction, some \safes support injections freely, consider
the simple example of the \emph{inner-r} string from \autoref{sec:safestrings}.
Admittedly this is a rather extreme example, but it serves to highlight the
following important point: in general it is not possible to know what actions over a
\safe are necessarily closed over that \safe. This means that the design of
every \safe needs to be bespoke and will support a different subset of possible operations.

This leads naturally to the question of error handling. There will be times when
an operation over a \safe does not result in a well-formed \safe. Depending on
requirements, there are two obvious ways to address this situation. One is to
throw an exception and fail eagerly. The other is to use an \lstinline+Either+
type, familiar from functional programming. \lstinline+Either+ is the commonly
accepted name for the coproduct, or disjoint union, of types in most functional languages.

\section{Realisation of \safes in \target}\label{sec:implementation}
To evaluate \safes, we implemented them in \target, because of its
powerful and flexible type system and support for both object-oriented and
function programming. Moreover, \emph{stringly typed} data is extremely
common in \target and \emph{JavaScript} code. \target already offers ways
to use literals, both string and number, as types (\autoref{sec:related}). Combined with type guards,
union types and other features of the type system, they give ample scope for
demonstrating the potential of \safes.

A \safe requires some way to model a structure and a way to have a recogniser over that structure.
One needs only to find an image of these constructs in the language to have
\safes. To model the structure in \target we
chose to use objects in which the fields of the object corresponded with
grammar productions. For the recognisers, we relied on \emph{parser combinators}.

\subsection{Parser Combinators}%
\label{subsec:parsercombinators}

We use parser combinators for our evaluation due to their easy composability.
\target has native support for \regexes, and indeed, at the lowest level, parser
combinators also make use of regular expressions. It is not clear what the full
expressive power of \regexes is in the language hierarchy, though they do have
limited ability to recognise context-free languages. Monadic parser combinators
are at least as expressive as this.

Regular expressions accept strings, but \safes require more than this.
Parser combinators do strictly more than accept; they can also process their
input. For \safes, this processing is the creation of an \AST.
We present a brief overview of parser combinators in the remainder of this
subsection. 

Parser combinators are familiar tools in functional programming, having been first introduced in 1975 by
Burge~\cite{burge} and popularised by many others, such as Hutton and Meijer~\cite{hutton92}.
They provide an elegant and declarative methods for implementing parsers. In contrast to traditional
parser generators, parser combinators are first class values which can be combined and manipulated
to define new combinators. They elide the difference between lexers and parsers.
They do not require a separate tool-chain, thus avoiding any
integration issues. A \safe library for \target therefore can be entirely
self-contained. It does not require any other tools to work apart from the
\target compiler itself.

A parser combinator is a self-contained unit. For example, one can write a
combinator \lstinline+oneOf+ that succeeds if the input character is an element
in its accepting list. Suppose the list is \lstinline+`` \t\n\r''+, then
\lstinline+oneOf `` \t\n\r''+ defines a parser that accepts exactly one whitespace.
To parser multiple whitespace, as in~\autoref{lst:spaces}, we combine, or compose, two combinators,
\lstinline+some+, which accepts one or more instances of another parser, and \lstinline+oneOf+.
\emph{Monadic} parser-combinators are able to parse context sensitive
grammars. This allows parser
combinators to handle XML documents in a single pass as in~\autoref{fig:xml}
(Leijen and Meijer~\cite{parsec-direct-style-monadic-parser-combinators-for-the-real-world}). 
This guarantees well-formedness in a single pass over the source.

\begin{lstlisting}[float=t,label=lst:spaces, caption=Parsing with combinator composition.]
spaces = some (oneOf " \t\n\r")
\end{lstlisting}

\begin{lstlisting}[float=t, label={fig:xml},%
caption={XML parsing in Haskell's Parsec library.}%
]
xml = do {
    name <- openTag
    content <- many xml
    endTag name
    return (Node name content)
    } <|> xmlText
\end{lstlisting}

There are some drawbacks to parser combinators. Parser generators give static termination guarantees
that most parser combinators cannot give. Work on total parser combinators, which provably terminate
is an area of active research
(Danielsson~\cite{Danielsson:2010:TPC:1932681.1863585}). So far at least, these
have suffered from poor performance.
Parser combinators are unable to deal directly with left-recursion. Most parser combinator libraries
provide a \lstinline+chain+ combinator (Fokker~\cite{fokker}) that captures the left-recursive design pattern without
obliging the user to rewrite a left recursive grammar to be right recursive (Aho \etal~\cite{Aho:86}).
There is a risk of inefficiency with parser combinators. However, this is not a necessary feature of
parser combinator libraries and many efficient implementations exist. \target
has at least one industrial strength library available\footnote{\url{https://github.com/jneen/parsimmon}}.

To model the grammar of a \safe we chose \target objects. This is primarily
for familiarity of programming model. Complex grammars might benefit from a more
abstract representation (\ie the visitor pattern used for an \AST), with purely
functional references in the form of lenses.
Finally, we also need a way to handle errors, as discussed in
\autoref{subsec:closure}. \target has good mechanisms for exception handling.

\subsection{Preprocessing for Natural Assignment}

We mention a preprocessing stage in \cref{sec:intro}. This processing stage is
very simple. Its sole function is to allow for declarations that are more
\emph{string-like} without needing to make changes to the \target itself. This
preprocessing is not required for \safes to function properly. The most common
usage is to desugar assignments (\cref{fig:desugar}). Preprocessing is also
required to allow for the usual \target operators on strings to work as
expected, if $x$ and $y$ are both \safes, then \lstinline+x == y+ is rewritten
as \lstinline+x.doubleEq(y);+.

\begin{lstlisting}[float=t,label={fig:desugar},caption={Desugaring in \target.}]
    // a SafeString assignment
    let file : FilePath = '/this/is/a/file.txt';
    // the preprocess changes this tool
    let file : FilePath = new FilePath('/this/is/a/file.txt');
\end{lstlisting}

The preprocess stage has one other important function. A statically declared
\safe (\ie one that is hard-coded) is converted into an object. Unfortunately,
\target does not statically check an object's constructor. As all of the
required information is present to check this declaration, for simplicity we
pass those declarations through \emph{node} is order to evaluate the
constructors. It would be easy, though unnecessary for demonstration purposes,
to incorporate the parse check directly into the compiler without having to
evaluate the file.

\subsection{Alternatives to Parser Combinators}
\label{sec:deploy}

\begin{lstlisting}[float=t, label={fig:namedcapture},%
caption={Named capture groups in JavaScript can be used in a manner similar to a
\safe for simple structures \cite{namedcapture}.}%
]
let re = /(?<year>\d{4})-(?<month>\d{2})-(?<day>\d{2})/u;
let result = re.exec('2015-01-02');
// result.groups.year === '2015';
// result.groups.month === '01';
// result.groups.day === '02';
\end{lstlisting}

Parser combinators are not a strict requirement of \safes. The expressive powers of \emph{pcre2} regular
expressions is sufficient to capture most of the likely use cases for
type safe strings. The chief advantages of using parser combinators are
readability and compositionality. JavaScript and \target, as of
2018, have \emph{named capture groups} for regular
expressions\cite{namedcapture} (\autoref{fig:namedcapture}). These can be made
to correspond with simple instances of \safes. Note that they do
not create a new \emph{type}, but can work within an object wrapper when users
wish to create an uncomplicated \safe.

The code in~\cref{fig:namedcapture} is very close to what we want from a simple
\safe. Indeed, it automatic maps named capture groups to fields in a new
\safe object, with the associated sub-membership tests the (sub) regular expressions.
However, compose these named groups is difficult and 
any change requires rechecking the entire
string, unlike with a fully-fledged \safe.

Declaring new \safes of greater complexity than this is clearly a greater burden
on the programmer. One needs to identify an
appropriate grammar, write recognisers for that grammar, write methods over the structure, and write a class which encapsulates
everything. Even for a string like an email, this is a non-trivial effort. For complex structures,
the code is likewise more complex. 
To aid deployability, the most common \safe types
should be supplied in the form of a library. This might include email strings, css fragments, address
formats and other forms of structured data. Library writers with unique string types
could write the necessary code themselves if they require that level of type safety.
\label{subseq:deploy}

\section{\safes in Action}\label{sec:casestudies}
Now that we have detailed the formation of \safes and discussed the mechanics of
instantiating them in \target, we look in more detail some concrete examples of
\safes, why they are a suitable target for making \emph{safe} and how they are
constructed. In turn we consider filepaths, css elements and finally email
addresses. 

\subsection{\safe Email Addresses}\label{subsec:email}
Email addresses are a common type of string in web programming. Ensuring that
email strings are well formed is a standard procedure. Extracting information
from an email string is also common, especially slicing out the \lstinline+name+
part of an email address. Email strings as string literals
(\autoref{sec:related}) are too inflexible: they have far too much variety of form
to capture in a simple literal type. Indeed, in the \regex-validated strings
pull request \cite{regexlit}, safety for email strings consumes much of the discussion.
Representing them as \monoidal or \regex-validated strings goes some
way to solving the problem, but fails to allow for operations and subtyping (\Cref{fig:subtyperegex}). 

Intuitively, a subtype relation over email strings would be flat, with
each provider (\ie gmail or hotmail) being a subtype of a generic email string. This generic email
should capture the `essence' (\ie the free structure) of a email string, without being tied to specific detail. 

Subtyping of email addresses is now the action of overriding the parser for one of the declared
fields. This produces a natural, intuitive, notion of type hierarchy. Turning an Email into a Gmail
means changing the domain parser to accept `gmail' specifically. Unlike with regular expressions,
there is no decidability issue here either.

Operations over email strings can now be defined in such as way that they behave in a rational, type
safe way. Normal string concatenation for example will produce a new string of type string, \ie
perform an upcast to string by default. For email addresses however, string concatenation could be
defined solely over the \lstinline+name+ field of the \safe. Now only the parser combinator
associated with that field need be called. If this succeeds then the string is still an email
string. It is not necessary to re-parse the entire string, as would be required by a \pcre
approach. \lstinline+"foo@gmail.com".append('bar')+ produces a string \lstinline+foobar@gmail.com : gmail+. 
In languages which support operator overloading, one could overload `+' so that the correct definition is chosen
automatically. Types can then be checked with the typechecker.

This approach allows for a much more fine-grained handling of type safety. For example, rather than
having \lstinline+user : string+ as a projection from a gmail string, the projection can be
\emph{tagged} with its origin by wrapping it an in object of its own (\autoref{fig:projections}).

\begin{lstlisting}[float=t, label={fig:projections},%
caption={Provenance tracking in \safes. While projections from the structure can
be typed as string, they can also have their own newtype wrapper. A name
extracted from an Gmail \safe can signal its provenance by having the type
\lstinline+GmailName+.}%
]
    class GmailName {
        gmailName : string;

        constructor(n:string) {
            this.gmailName = parse(n);
        }
    }
    /* etc */

    class Gmail {
        name : GmailName
        domain : Domain
    }
    /* etc */
\end{lstlisting}

Now the different fields of an email address can signal their origin even when detached from the
original object. They signal the provenance of a string and reveal that it has
been detached from another structure. This can be exposed to the typechecker, or
treated as a simple type alias
(\autoref{sec:related}). This latter serves as useful documentation to the user.
With the former, it is possible to define operations over these projections in much the
same manner as for strings of type Gmail. Such levels of type safety are not perhaps required for
many simple programming tests, but the approach allows for it when required, rather than insisting
on it.

\subsection{\safe FilePaths}\label{subsec:filepath}
\begin{lstlisting}[float=t,label={lst:windowspath},framerule=0pt,mathescape,
caption={A \regex-validated WindowsPath, where \textbackslash w is any alphanumeric
character.}
]
  type WindowsPath = ^((((([a-z]|[A-Z])[:\|])?(\\((\w|(\w[ \.]?\w))+(\\(\w|(\w[ \.]?\w))*)*)?))|
    (\\?(((\w|(\w[ \.]?\w)))+(\\(\w|(\w[ \.]?\w)))+(\\(\w|(\w[ \.]?\w))*)*)?))
    (\\(\w|(\w[ \.]?\w))+\.?[^\.]+)?(\:(\w|(\w[ \.]?\w))+)?)$\mbox{\textdollar}$
\end{lstlisting}

Filepaths are a common feature in many programs and are usually typed as \str.
Some languages (\eg Haskell) introduce a \lstinline+FilePath+ type that is, in
fact, merely a type alias of \str. Filepaths have different formats on different
architectures, some using \lstinline{\}, some \lstinline{/}, or even a mixture
of the two. Manipulating filepaths in the shell is a well known problem, relying
on \lstinline+basename+ and \lstinline+realpath+ to extract information.
The shell is a unityped environment: everything is a string. While shells could
benefit from the versatility of \safes, the
problem exists equally in strongly typed programming languages.

A \monoidal \safe is the simplest way to gain increased type safety for
filepaths.  \autoref{lst:windowspath} contains a \regex-validated \emph{Windows}
pathfiles.  It is immediately clear that this is not a useful constraint in a
type. It is completely unmodular. Worse, it does not accept ``/'' as a directory
separator, though this is valid for \emph{MicroSoft Windows}.  Closer inspection
reveals that the \regex contains a great deal of repetition. This repetition can
be factored out when using parser combinators, and opens up the possibility of
overriding one or more of the substrings to create specialised parsers
(\autoref{subsec:subtyping}). Using a monadic parser combinator also solves the
problem of consistent directory separators. The monadic state can be updated to
accept whichever separator comes first and then propagate that constraint over
the rest of the string.

Assuming that we do not want to use a `simple' \monoidal \safe, 
one possible representation of a filepath might be:
\begin{lstlisting}[framerule=0pt]
  (FilePath (Path array<Directory>) (File (Name string) (Ext string)))
\end{lstlisting}
where we do not need to represent the separator nor the ``.'' in the filename
proper as they are invariants. This representation assumes that we wish to normalise the separator over
the entire program, \ie \lstinline+cast()+ would be:
\begin{lstlisting}[framerule=0pt,mathescape=true]
  cast() : string = dirs.join('$\textbackslash$') + name.cast();
\end{lstlisting}
but there is no reason why we cannot also record the direction of the file
separator. This could be used, amongst other aspects of the string, to induce a subtype
relation for Filepaths based on separator direction, making a FilePath \safe generic over \emph{Unix} and
\emph{Windows} style paths. \emph{UnixPath} and \emph{WindowsPath} are now
subtypes of FilePath, which in turn is still fully interoperable with \str
(\Cref{lst:sep}). This is conditioned solely on the granularity of the
recogniser and representation.

\begin{lstlisting}[float=t,caption={Subtyping for filenames via the use of the
    separator.},label={lst:sep}]
  class UnixPath extends FilePath {
    separator = '/';
  }
\end{lstlisting}

The internal representation of a \emph{FilePath} contains a traversable structure, the
array of directories. Directories can be represented by raw strings (\ie without
associated recognisers, though further type safety could be had by providing checks). 
Extracting filename, extension and path are just named projections from the
structure.
Type safety can be taken further still by putting constraints over
the length and contents of \lstinline+array<Directory>+
(\autoref{fig:bash}). The location of a file can be encoded in the type
information and checked dynamically.

Suppose a shell with support for \safes with method overloading.
This is not unreasonable, as \safes can function as simple strings when required.
In it, the shell would allow you to define a \safe \lstinline+HomePath+. A
string of type \lstinline+HomePath+ points to a file in the home directory.
overload \lstinline+mv+ temporarily to mv so that it only functions from
HomePath to one of its descendants. One can then call \lstinline+mv+ safe in the
knowledge that in incorrect assignment is now impossible. 

\begin{lstlisting}[language=bash,float=t, label={fig:bash},%
caption={Scripting in a \safe-aware shell environment could be made much safer.
  A type safe \emph{rm} would make it impossible to delete files in a sensitive
  directory or of a particular type, but the shell itself could remain essentially \emph{unityped} with
  everything being considered a string.}%
]
    # HomeDotFile is a SafeString where the file must be located in the home
    # directory and be hidden. 
    HomedotFile = { dirs : [home], filename = /\.[chars]/}
    # temporarily overload rm so that it cannot delete HomeDotFile strings.
    rm ~/.bashrc # fails as ~/.bashrc : HomeDotFile
  \end{lstlisting}

\subsection{\safes in CSS and Web-forms}\label{subsec:css}
Web-based programming is especially rich in strings.  Cascading Style sheets
(\css) is an almost ubiquitous style sheet language that is fundamentally
stringly-typed. This makes it very easy to incorrectly assign data with little
that either a type checker or static analyser can do to prevent it.  
Here we show how \safes solve this problem using four examples:  \css colours,
phone numbers, units, and cross-site scripting.

\boldpara{CSS Colours}
We have already see in \autoref{sec:intro} some of the problems 
caused by representing colours as strings. 
Analysing a \cssColour \str reveals an underlying structure despite the
superficial differences in presentation. Both \lstinline{``#000''} and
\lstinline{``#000000''} encode the same information, but unimportant differences
restrict their interoperability. Even simple equality fails:
\lstinline{``#000''} does \emph{not} equal \lstinline{``#000000''} even though
they represent the same colour.  Abstracting a \cssColour \str, however, gives
the structure 
\lstinline[mathescape]+(Colour (Red $\mathbb{N}$) (Green $\mathbb{N}$) (Blue $\mathbb{N}$))+
which elides these superficial differences.  Regular expressions can instantiate
this structure for a particular string, but it is accomplished even more easily
with a \emph{parser}:
\begin{lstlisting}
cssColour = parseHash.then(parseRed.then(parseGreen.then(parseBlue)))
\end{lstlisting}

If the parser accepts, then the raw string is a \cssColour and can be
represented as a record (\autoref{lst:cssColObj}).
\begin{lstlisting}[float=t,mathescape,caption={A concrete realisation of AST$^{PL}$ for a cssColour},label={lst:cssColObj} ]
  class CssColour {
    red   : number
    green : number
    blue  : number
  }
\end{lstlisting}
Equality is now optionally over the structure of the \cssColour.  The original
strings can, of course, still be compared \emph{as strings}, using \safes' 
\lstinline+cast()+ operation.  For a \cssColour the definition is shown in \autoref{lst:castcss}.
The '\#' symbol is \emph{invariant}, so it does not need to be stored.

\begin{lstlisting}[float=t,caption={CSS colour string \lstinline+cast+.},label={lst:castcss}]
  cast() : string = '#' + red.toString(16) + green.toString(16) + blue.toString(16)
\end{lstlisting}

This \lstinline+CssColour+ representation, and the fact that 
it contains only valid hexadecimal colour strings, is much
easier to manipulate than raw strings. Changing elements \emph{in} the structure
leaves the structure itself, notably its type, unaltered. For example, say we 
wanted to blend two colours together. This
is not easily done when dealing with raw strings:
\begin{lstlisting}
  '#123' + '#332211' // naive: '#123#332211', an obviously invalid cssColour
\end{lstlisting}

\begin{lstlisting}[float=t,framerule=0pt,mathescape,
 caption={Regex for US phone numbers taken from from a \emph{stackoverflow} answer.},
 label={lst:phone:regex}
]
  ^(?:(?:\+?1\s*(?:[.-]\s*)?)?(?:\
  (\s*([2-9]1[02-9]|[2-9][02-8]1|[2-9][02-8][02-9])\s*\)|
  ([2-9]1[02-9]|[2-9][02-8]1|[2-9][02-8][02-9]))\s*
  (?:[.-]\s*)?)?([2-9]1[02-9]|[2-9][02-9]1|[2-9][02-9]{2})
  \s*(?:[.-]\s*)?([0-9]{4})(?:\s*(?:#|x\.?|ext\.?|extension)\s*(\d+))?$\mbox{\textdollar}$
\end{lstlisting}

\begin{lstlisting}[float=t, label={fig:phone},%
caption={A part of the realisation of a US phone number \safe in \target, using the
  'parsimmon' parser combinator library. This code fails with a parse error if
  the string is not accepted.}
]
  import * as P from 'parsimmon'

  let separator =
      P.oneOf(" \./-").fallback('');

  let areacode = P.regexp(/[2-9][0-9]{2}/).map(Number).desc('a valid area code');
  let nxx = P.regexp(/[2-9][0-9]{2}/).map(Number).desc('a valid office code');
  let xxxx = P.regexp(/[0-9]{4}/).map(Number).desc('a valid identifier');

  let phoneParser =
      P.seqMap(
          areacode,
          separator,
          nxx,
          separator,
          xxxx,
          function(a, s1, n, s2, x) { return [a, n, x] }
      );

  class USPhone {
      area : number;
      office : number;
      uniq : number;

      constructor(phone : string) {
          let pn : number[] = phoneParser.tryParse(phone);
          this.area = pn[0];
          this.office = pn[1];
          this.uniq = pn[2];
      }

      cast() : string {
          return this.area.toString() + "-" + this.office.toString() + "-" + this.uniq.toString();
      }
  }
\end{lstlisting}

\boldpara{Phone Numbers}
Phone numbers entered on websites make excellent \safes.
Unfortunately, there is no universal model for phone number
construction to act as a supertype from which other formats can be derived
(\cite{wiki:phone}).
Obtaining a \regex to correctly match US phone numbers is difficult. Consider
\autoref{lst:phone:regex}, which shows a regex taken 
from a \emph{stackoverflow} answer \cite{phonenumber}.
One can readily see that this is inappropriate for a type constraint, and highly
inflexible. Underlying this surface complexity, there is the \emph{NANP}: North
American Numbering Plan \cite{nanpa}.
This gives a formatting convention of \lstinline+NPA-NXX-XXXX+ where
\lstinline+NPA+ is a three digit area code and the remaining seven digits are
the subscriber number. This maps neatly into the structure in
\lstinline+(NPA+ $\mathbb{N}$ \lstinline+(Central+ $\mathbb{N}$\lstinline+)+
\lstinline+(Ident+ $\mathbb{N}$\lstinline+))+
where \lstinline+N+ is a digit from 2--9.  Whitespace, dots, dashes, or some
combination may separate the groups. Parser combinators easily parse a number
string into this structure, even when a mix of separators is used.
Appropriately defining \lstinline+cast()+ cleanly normalises the format, without
resorting to functions such as \lstinline+replaceAll()+. 

Parsers expect structured data: the code in \cref{fig:phone} is reasonably
generous in what it accepts. It recognises as a phone number 
\lstinline+555-211 1234'+ with mixed separators, and \lstinline+5552111234+ 
as valid phone numbers.
It does not accept other groupings such as \lstinline+55 52111 234+. The
well-known \emph{Robustness Principle} or \emph{Postel's Law} \cite{postel}
states \say{Be liberal in what you accept, and conservative in what you send}
but it is also well-known that this maxim has harmful
consequences~\cite{thompson}. It is a difficult design decision to know how
flexible to be when accepting input: in this case, we require that the input
have at least the structural grouping of a phone number.

\begin{lstlisting}[language=html,float=t,label={lst:cssUnit},
caption={An easy typing error leads to a difficult to
    find mistake in \css. Linting tools \emph{may} find the problem, but as a
    \safe this unit error would be caught automatically.}
]
  <!-- we assume that margin, width, padding etc have type (PX | AUTO) in this example --!>
  object { height : 250px; width : 200px; margin: 10pc; margin-bottom: 10px;
    padding: 5px;}
\end{lstlisting}

\begin{lstlisting}[float=t,caption={The complexity of adding incompatible units can be
    abstracted away from the front-end user.}, label={lst:pixels}]
  // overloading + for a pixel string
  add(operand : Pixel | Picas | Points) {
    if (operand isinstanceof Pixel) { return cm self.value + operand.value; }
    else if (operand isinstanceof Picas) { return cm self.value + operand.toCm(); }
    // ... etc
  }
\end{lstlisting}

\boldpara{Units} Units in \css, such as \emph{pixels}, \emph{points} and
\emph{picas}, are all stringly typed.  This makes it extremely easy to assign
the wrong unit to the wrong variable. It also makes it very difficult to a
static analysis to identify these errors. Yet, the latent structure of these
strings is scarcely latent at all.  Fortunately, \safes easily solve the latent
structure problem for these strings.

A string ``1cm'' has the simple
representation \lstinline+(CM 1)+. It is trivial to catch the incorrect
assignment \lstinline+let font-size : Point = '1cm'+. It may seem that modelling
these strings as \safes is excessive. This is not the case.
\autoref{lst:cssUnit} has a simple mistake, $c$ was typed instead of $x$ in the
unit for \lstinline+margin+; spotting this error in a dense page of \css is
difficult and time-consuming. A linter may not detect this as an error. A
declaration that \lstinline+margin+ takes the union type \lstinline+(PX | AUTO)+
catches the error without programmer effort.

Stripping the unit information and normalising the numerical representation also
makes it much easier to define conversion functions that do not expose their
complexity to the front-end user. This can be achieved by, for example, operator
overloading. If operator overloading is not desired (\target for example does
not encourage it), then \target's union types can be used to overload a function
to choose the correct transformation (\cref{lst:pixels}).

\begin{lstlisting}[float=t,language=html, label={fig:xss},%
caption={A simple example of XSS scripting in HTML as found in 
\cite{chen2017decidable}.  Typing user input with
  \safes significantly reduces the surface area for XSS attacks. Moreover, the
  structure of the user input is now known and can be used elsewhere in the code
  without having to recheck it. The \safe Sanitised acts as a certificate for
  the contents of the input string; UserName does the same.}%
]
    <!-- unsafe html --!>
    <h1> User <span onMouseOver=''popupText('{{bio}}')''>{{userName}}</span></h1>
    <!-- the same code, with \safes for added safety --!>
    <h1> User <span onMouseOver=''popupText('{{bio : Sanitised}}')''>{{userName : UserName}}</span></h1>
\end{lstlisting}

\boldpara{Cross-site Scripting}
Poor control of strings can
result in bugs that are merely irritating, \ie layout and colour problems, or
bugs that are dangerous.
XSS (cross-site scripting) falls into the latter category. The \lstinline+replaceAll+
function is often used to sanitise user input in an attempt to mitigate the
danger of XSS attacks \cite{chen2017decidable}. They take as a running example
the code fragment in \autoref{fig:xss}. A \safe for sanitised data could be a
\monoidal \safe, habitation of the type being a proof that the input string does
not contain a code injection. Now the string is \lstinline+Sanitised+ \safe.
Such a string can be used elsewhere without being reexamined, amortising the initial cost of checking.
Such a check would be performed anyway in conscientious code, but knowledge of
the sanitisation would not be passed to the type checker, leaving the programmer
to check the string again and again.

\begin{lstlisting}[float=t,caption={\safes for SQL queries can substantially
    reduce the surface for an injection attack.},label={lst:sql}]
  // a typical SQL query with a vulnerability
  statement = `` SELECT * FROM users WHERE name = ' '' + userName `` ';'' 
  // if userName = ' OR '1'='1' --  (bad: a SQL injection)
  // an appropriate UserName sanitises the string, but also passes the
  // information to the type checker
  statement = `` SELECT * FROM users WHERE name = ' '' + userName : UserName `` ';'' 
\end{lstlisting}

SQL injection attacks can also be reduced via \safes. \autoref{lst:sql} contains
a simple SQL injection. Defining a \lstinline+UserName+ \safe can effectively
sanitise the user input. Escaping via \eg \lstinline+replaceAll()+ is error prone, but
abstracting dangerous elements of the input string into an \AST and then using a
\lstinline+cast()+ operation that cannot produce malformed input is a much safer
approach.

\section{Related Work}\label{sec:related}
As there is little theoretical that
corresponds directly to our approach with \safes, we first
focus on work that we believe is compatible, such as the use of
constraint solvers, symbolic execution, the use of dependent and
liquid types, and \regex-validated string types. We go on to
compare \safes with various, mostly informal, approaches to the
string latent structure problem such as type aliasing, the use of
new types, object wrappers, and string literal types.

\subsection{Liquid Types and Dependent Types}

Type systems of differing levels of granularity and expressiveness abound both
in the literal and in practice. \safes solution to the latent structure problem 
has a strong overlap with dependent and liquid types. \safes are programmable in a
dependently typed language and offer very strong guarantees:
under the \emph{Curry-Howard}
correspondence \cite{Sorensen}, types are propositions and functions proofs. Making proofs out
of functions is powerful, but requires much from a programmer
intent on delivering a product. The code in \autoref{lst:agda} allows the user
to pass around not merely the structure itself, but a computer verified proof
that the string is in the given language. Most languages do not offer this
possibility, but not having a proof to consult does not mean that safety is compromised.

\begin{lstlisting}[float=t,language=Haskell, mathescape=true,
caption={What price safety? Dependent types allow a \safe to carry with it its
  own proof of correctness. The programming burden for this style is non-trivial.}, label={lst:agda}]
record SString : Set where
    constructor mkString
    field
        s : String -- the raw string
        re : RegEx -- the recogniser
        p : SubString re (unpackString s) // this is a proof of membership

makeSafe : String $\rightarrow$ RegEx $\rightarrow$ Maybe SString
makeSafe s re with s $\in$? re
makeSafe s re | yes pr = just $\mbox{\textdollar}$ record {s = s; re = re; p = pr }
makeSafe s re | no x = nothing

helloString : SString
helloString = fromJust $\mbox{\textdollar}$ makeSafe "hello" hello 
\end{lstlisting}

The goal of both dependent types and liquid types is to make the type level
abstraction of program behaviour more expressive and less coarse. 
Strings have seen little research in this area, perhaps due to the problems of
decidability in string theory \cite{ganesh2011decidable}.

\subsection{Regular Expression Validation}

Regex-validated string types of~\cite{regexlit} allow for a high degree
of type safety \wrt to strings. \safes are strictly more expressive in that they
are not tied to one form of membership test (\ie PCRE2). Arbitrary logic can be
included in the recogniser for a \safe.
\safes also support subtyping, and operations
over \safes. These operations can be overloaded to support \emph{ad hoc}
polymorphism. The maintenance burden of \safes is not much more that
\regex-validated types. For simple declarations, the newly introduced \emph{named
  capture groups} can stand in the stead of parser combinators
(\autoref{sec:deploy}).
Indeed, there is no theoretical necessity for parser combinators in the \target implementation
except for the noted provisos in~\autoref{sec:deploy} of subtyping and readability.

\subsection{Industrial Approaches}

There are a number of informal approaches to increased string type safety which
make for reasonably usable design patterns.

The simplest is \emph{type aliasing} \cite{advanced}. This is supported in many languages, but
its primary use case is often for documentation purposes rather than type safety.
In \target, for example, one writes \lstinline+type Name = string+.
From the perspective of enforcing invariants however, the utility of aliasing is
limited because it cannot enforce type (in-)equality (\autoref{lst:alias}).
It cannot prevent static type errors of the kind \lstinline{gmail : Gmail = "not a gmail address"},
where \lstinline+Gmail+ is some previously declared type alias for \str.

\begin{lstlisting}[float=t,caption={Type aliasing in \target},label={lst:alias}]
  type Name = string;
  foo : Name = 'hello';

  bar : string = 'hello';

  foo == bar -> True // we probably want this to be false
\end{lstlisting}

\begin{lstlisting}[float=t,caption={\target Inferfaces can be used to create \emph{new
      types} by wrapping other types.}, label={lst:newtypes}]
  interface Name { name : string }
  let foo : Name =  { name 'hello' }
  foo == 'hello' // results in a 'no overlap' error, rather than just false
\end{lstlisting}

If a language has \emph{newtypes}, greater type safety can be achieved. These
are a wrapper around an already existing type. A \emph{newtype} is different
from am alias in that it is exposed to the type checker and can be used
to enforce inequality.
Object-orientated languages support these by wrapping a
primitive type in an object or using an interface \cite{advanced}.
\autoref{lst:newtypes} shows the syntax in \target.
This object wrapper pattern can be validated with a recogniser
(\autoref{lst:reObj}). A simple instance of
this pattern is explored in \cite{fowler}. The constructor is augmented with a
\regex check for string membership. The constructor fails with an error if the string is not in the
language expressed by the \regex. This pattern is captured by \monoidal \safes.

\phantomsection\label{subsec:stringlits}
Since version 1.8, \target has had \emph{string} and \emph{number} literal types
\cite{advanced}. The original pull request for string literals~\cite{stringlit} summarises them as: 
``A string literal type is a type whose expected value is a string with textual contents equal to
that of the string literal type.''
A string literal type can only be assigned the exact value specified in that type~\autoref{lst:stringlit1}.
String literal types can be used with other features of the type system, most notably \emph{union
types}, to create (finite) enumerable sets of strings. This allows them to act
as type guards in pattern-matching in a similar way to type constructors in
pattern-matching within the ML family (\autoref{lst:stringlit2}).
\emph{Regex}-extended string literals relax the need for exact matching,
making string literal types even more expressive.

\begin{lstlisting}[float=t,label=lst:stringlit1, caption={A simple example of
    string literals in \target to increase type safety for functions which take
    strings as parameters.}]
const good :  "click" = "click";; // this is OK 
const bad  : "click" = "notAClick" // this assignment fails
\end{lstlisting}

\begin{lstlisting}[float=t,%
caption={A suggested use for string literals, as proposed in the original pull
request \cite{stringlit}. This gives considerable expressive power and safety to 
functions that take strings. \safes make this even more expressive.},
label={lst:stringlit2}]
type Direction = "North" | "South" | "East" | "West"
function move(direction : Direction) { 
    if (direction == "North") {
        // move north
    } .... // enumerate the options
}
move("SouthWest") // error: not one of our directions
\end{lstlisting}

\begin{lstlisting}[float=t,language=TypeScript, caption={An object wrapper with a
    membership check. This pattern corresponds to \monoidal \safes.}, label={lst:reObj}]
    class Name {
        name:string;

        /* optional regex check for membership */
        constructor(name:string) {
            this.name = name
        }
    }
\end{lstlisting}

The approach in \autoref{lst:reObj} is a design pattern
which can be implemented in any language with records and access to regular
expressions \cite{fowler}. This gives the pattern a
distinct advantage over \regex-validated strings and string literals. It does
not require any changes to the target language to introduce these features. It
uses only what the language already has.
This is useful from the point of view of safety, but is
problematic in some respects. Most obviously, it lacks the flexibility of
\safes; making operations more difficult to define, and limiting subtyping to
the nominal system.

\section{Conclusion}\label{sec:conclusion}
We have presented \safes, a language-agnostic approach to type safety for strings.
\safes introduce many of the benefits associated
with liquid types into an area previously little explored. \safes require no special
language mechanisms, requiring only an ability to encode structures and recognisers. One need
only find the image of these concepts within the language to encode a \safe. \safes are applicable
either statically or dynamically. Indeed, \safes squeeze more invariants out of simple type systems 
by treating strings as algebraic structures. The complexity of advanced language
features, such as dependent types and GADTs, are thus not required. A simple
type system can go further by putting the burden of representation on the
producer of library code, rather than the consumer.

We have also presented an instantiation of \safes in \target, showing the relative ease
with which they can be encoded and used. Such strings represent a fragment of what can be achieved with
dependent types in a non-dependently typed system. We do this by translating a structure over which we
have no particular control (\ie \str) into a form that exposes rich latent
structure to the type checker. All
artefacts are available at \url{anonymised.repo.to.do}.



\medskip
\bibliographystyle{ACM-Reference-Format}
\bibliography{references}

\end{document}